\documentclass[aps,prl,twocolumn,superscriptaddress,nofootinbib]{revtex4-1}

\usepackage[pdftex,colorlinks]{hyperref}
\usepackage[dvipsnames]{xcolor}
\definecolor{phthaloblue}{rgb}{0.0, 0.06, 0.54}

\hypersetup{
    colorlinks=MidnightBlue,
    linkcolor=MidnightBlue,
    citecolor=MidnightBlue,
    filecolor=MidnightBlue,
    urlcolor=MidnightBlue,
    }
\usepackage[pdftex]{graphicx}
\usepackage{multirow} 
\usepackage{bm}
\usepackage{slashed}
\usepackage{cancel}
\usepackage{here}
\usepackage{comment,braket}
\usepackage{epsf}
\usepackage{amsmath}
\usepackage{graphics}
\usepackage{amsfonts}
\usepackage{amssymb}
\usepackage{latexsym}
\usepackage{color}
\usepackage{natbib}
\usepackage[normalem]{ulem}
\input{colordvi.tex}
\usepackage{feynmp}
\usepackage[utf8]{inputenc}
\DeclareUnicodeCharacter{2212}{-}
\usepackage{orcidlink}
\begin{document}

\title{
Alignment and Enhanced Multi-Higgs Production 
}

\author{Subhojit Roy\,\orcidlink{0000-0001-6434-5268}}
\email{sroy@anl.gov}
\affiliation{High-Energy Physics Division, Argonne National Laboratory, Argonne, IL 60439, USA}

\author{Carlos E.~M.~Wagner\,\orcidlink{0000-0001-6407-623X}}
\email{cwagner@uchicago.edu}
\affiliation{High-Energy Physics Division, Argonne National Laboratory, Argonne, IL 60439, USA}
\affiliation{Enrico Fermi Institute, Department of Physics, University of Chicago, Chicago, IL 60637, USA}
\affiliation{Kavli Institute for Cosmological Physics, University of Chicago, Chicago, IL 60637, USA}
\affiliation{Leinweber Center for Theoretical Physics, University of Chicago, Chicago, IL 60637, USA}
\affiliation{Perimeter Institute for Theoretical Physics, Waterloo, Ontario N2L 2Y5, Canada}

\date{\today}

\begin{abstract}
Contrary to conventional expectations, we identify a class of extended scalar-sector scenarios in which final states with two, three, or four Higgs bosons constitute the leading discovery channels for new physics at the LHC. In these scenarios, higher-dimensional interactions, together with suppressed Higgs–scalar mixing near the alignment limit, reorganize the decay patterns of new scalar states, suppressing conventional modes while enhancing multi-Higgs final states. We illustrate the emergence of dominant triple- and quadruple-Higgs signatures in two representative realizations: a single-scalar extension of the Standard Model, where higher-dimensional operators suppress conventional two-body decays while preserving couplings to higher-multiplicity Higgs final states; and a two-singlet scenario, where similar signatures arise through cascade decays with a simpler operator structure. In both cases, the new scalar states can be produced via gluon fusion, 
yielding potentially observable rates for multi-Higgs production at the LHC.
Although both realizations lead to identical final states, they exhibit distinct kinematic features reflecting their underlying topologies, providing a direct handle on the dynamics.
\end{abstract}

\maketitle

\section{Introduction}

The discovery of a Higgs boson ($h$) with mass near $125$~GeV~\cite{ATLAS:2012yve,CMS:2012qbp} completed the particle content of the Standard Model (SM) and initiated a program to directly probe the structure of the scalar sector~\cite{Cepeda:2019klc, DiMicco:2019ngk}. A central objective is to uncover the form of the Higgs potential, which remains only indirectly constrained. Current collider strategies focus on Higgs-pair production as the leading probe of the scalar potential, providing direct sensitivity to the Higgs trilinear self-coupling, while higher-multiplicity Higgs final states offer access to quartic interactions. Resonant searches for physics beyond the SM (BSM) further extend this program by targeting new scalar states that can enhance these processes and provide complementary probes of extended scalar sectors. Future collider programs are expected to significantly extend the sensitivity to Higgs self-interactions and to the structure of extended scalar sectors~\cite{AlAli:2021let, deBlas:2022aow,Forslund:2023reu, DiVita:2017vrr,Tian:2013yda,Barklow:2017suo,dEnterria:2017dac,Forslund:2022xjq,Forslund:2023reu,Li:2024joa,Maltoni:2024dpn}.

At the LHC, extensive efforts have focused on both~resonant and nonresonant di-Higgs production~\cite{ATLAS:2024ish,CMS:2022cpr,DiMicco:2019ngk,Cepeda:2019klc,ATLAS:2023vdy,CMS:2024phk,CMS:2022gjd,CMS:2020tkr,CMS:2026nuu,Collaboration:2928096,CMS:2024pjq,ATLAS:2019qdc,ATLAS:2022xzm,ATLAS:2018uni,ATLAS:2022hwc,ATLAS:2021ifb,ATLAS:2025eii,CMS:2025qit,CMS:2017rpp}. More recently, the experimental program has begun to extend beyond di-Higgs signatures, with dedicated searches for triple-Higgs production, including resonant cascade topologies~\cite{CMS-PAS-HIG-24-012,ATLAS:2024xcs, CMS-PAS-HIG-24-015, Abouabid:2024gms}. While these studies already probe part of the relevant signal space, they are not yet guided by a framework in which multi-Higgs final states constitute the dominant collider signatures.

This situation reflects the prevailing theoretical expectation that, in most motivated scenarios, $SU(2)$ symmetry together with the Goldstone Equivalence Theorem relate the Higgs and longitudinal gauge-bosons ($W,Z$) final states, leading to correlated decay patterns of new scalar states across the $hh$, $WW$, and $ZZ$ channels~\cite{Cornwall:1974km,Lee:1977eg,Vayonakis:1976vz,Chanowitz:1985hj,Veltman:1989ud}. In particular, mixing between a new scalar and the SM-like Higgs induces couplings to all SM fields, including fermions ($f$), such that the decay rates of the new scalar into $hh$, $WW$, $ZZ$, and $f \bar f$ are controlled primarily by the same mixing angle and are therefore correlated. As a result, di-Higgs signatures are generally accompanied by observable signals in other SM final states. In addition, higher-multiplicity Higgs final states, such as $hhh$ and $hhhh$, are suppressed by phase space and by the structure of scalar interactions, rendering them subleading compared to two-body channels. Together, these considerations underpin the standard expectation that the dominant collider manifestations of extended scalar sectors arise in low-multiplicity scalar final states.

These expectations, however, need not hold. Extended scalar sectors ($S$) can realize regimes in which the decay hierarchy of new scalar states is reorganized, such that higher-multiplicity Higgs final states dominate over conventional two-body decays into SM states. In particular, the partial widths can satisfy
\begin{equation}
\Gamma(S \to WW,\, ZZ,\, \bar f f) \ll \Gamma(S \to hh),
\end{equation}
or even
\begin{equation}
\Gamma(S \to hh,\, WW,\, ZZ,\, \bar f f) \ll \Gamma(S \to hhh),\;\Gamma(S \to hhhh),
\end{equation}
leading to scenarios in which multi-Higgs final states become the primary collider signatures. This hierarchy arises from the interplay of alignment and higher-dimensional interactions, where the former suppresses couplings to SM fields, while the latter allows sizable scalar trilinear interactions to persist. Even in the alignment limit, where decays into gauge bosons and fermions are strongly suppressed, the $S \to hh$ channel can remain present, but can also be parametrically reduced relative to higher-multiplicity Higgs final states.

In this Letter, we identify a class of scenarios in which resonant multi-Higgs final states emerge as the dominant collider signatures of extended scalar sectors. This behavior can arise through two complementary mechanisms: direct multi-Higgs decays of a single scalar and cascade processes in two-scalar setups~\cite{Papaefstathiou:2020lyp,Papaefstathiou:2025meh}. The corresponding collider signatures can be summarized schematically as
\begin{equation}
pp \;\to\;
\begin{cases}
\text{cascade:} & s_1 \to h s_2 \;\; (s_2 \to hh) \;\to hhh, \\[4pt]
                & s_1 \to s_2 s_2 \;\; (s_2 \to hh) \;\to hhhh , \\[6pt]
 \text{direct:} & s \to hh,\; hhh,\; hhhh \, .
\end{cases}
\end{equation}
Identical final states, $hhh$ and $hhhh$, can arise from structurally distinct underlying dynamics. Despite this, the two mechanisms lead to distinct kinematic features, providing a direct handle for experimental discrimination at colliders. In both realizations, the heavy scalar states ($s_1$, $s_2$ or $s$) can be produced via gluon fusion, e.g.\ through vector-like quark (VLQ) loops, while their decays generate the multi-Higgs final states, defining a regime in which $hh$, $hhh$ and $hhhh$ production provides the leading experimental handle on the underlying dynamics.

This regime contrasts with most previous studies, where enhanced multi-Higgs production is typically accompanied by correlated signatures in gauge boson and fermionic channels~\cite{ATLAS:2020fry,CMS:2019kaf}. In the scenarios identified here, these conventional decay modes are parametrically suppressed, leaving multi-Higgs final states as the leading probes of the extended scalar sector and motivating a shift in experimental search strategies.

To make these features explicit, we now turn to the underlying framework and its collider implications. We first present the theoretical setup, including both the two-singlet cascade scenario and the one-singlet effective realization, and show how the interplay of alignment and higher-dimensional interactions leads to dominant multi-Higgs final states. We then discuss the collider implications, emphasizing the kinematic structure of the signal and its experimental interpretation. Technical details are discussed in the Appendices.
\section{Model setup and mechanism}
\label{sec:model}
\subsection{Two-real singlet scalars extension}
We extend the SM by two real gauge-singlet scalars, $S_1$ and $S_2$, and a VLQ $T\sim(3,1,2/3)$. After electroweak symmetry breaking,
\begin{equation}
H=
\begin{pmatrix}
G^+\\[2pt]
\dfrac{v+h+iG^0}{\sqrt2}
\end{pmatrix},
\qquad
S_1=v_1+s_1,
\qquad
S_2=v_2+s_2,
\end{equation}
with $v\simeq246$ GeV. The relevant Lagrangian is
\begin{align}
\mathcal L \supset &\ \frac12(\partial_\mu S_1)^2+\frac12(\partial_\mu S_2)^2
-\ V(H,S_1,S_2) \nonumber\\
&+ \bar T(i\!\not\!\!D-M_T)T-y_{T,1} s_1 \bar T T -y_{T,2} s_2 \bar T T\,.
\label{eq:Lag}
\end{align}
The VLQ induces a loop-generated coupling of the heavier singlet-like state to gluons~\cite{Chen:2017hak}. In the heavy-fermion limit $M_T\gg m_{s_1}, m_{s_2}$, integrating out $T$ gives
\begin{equation}
\label{s1ggeff_main}
\mathcal L_{\rm eff}\supset
\frac{\alpha_s}{12\pi}\frac{y_{T,1}}{M_T}\,
s_1\,G_{\mu\nu}^aG^{a\mu\nu} + \frac{\alpha_s}{12\pi}\frac{y_{T,2}}{M_T}\,
s_2\,G_{\mu\nu}^aG^{a\mu\nu}
\end{equation}
which enables resonant $s_1$ and $s_2$ production through gluon fusion at the LHC. We take $M_T\sim {\cal O}(\text{TeV})$, and take Yukawa couplings so that current direct limits can be satisfied while retaining an appreciable production rate.
At the renormalizable level, the phenomenologically motivated scalar potential can be parametrized~as,
\begin{align}
V_{\rm ren}=&-\mu_H^2 H^\dagger H+\lambda_H(H^\dagger H)^2
+\frac{m_1^2}{2}S_1^2+\frac{ m_2^2}{2}S_2^2 \nonumber\\
& +\mu_{12}^2S_1S_2 +a_1S_1H^\dagger H+a_2S_2H^\dagger H \nonumber\\
& +\kappa S_1S_2H^\dagger H
+\frac{\lambda_{H S_1}}{2}S_1^2H^\dagger H
 +\frac{\lambda_{H S_2}}{2}S_2^2H^\dagger H
\nonumber\\
&
+\frac{\lambda_1}{4}S_1^4+\frac{\lambda_2}{4}S_2^4
+\frac{\lambda_{12}}{2}S_1^2S_2^2,
\label{eq:Vren}
\end{align}
supplemented by the leading higher-dimensional term
\begin{equation}
V_{\rm dim5}=\frac{c_5}{\Lambda}S_2(H^\dagger H)^2,
\label{eq:Vdim5}
\end{equation}
where $V(H,S_1,S_2) = V_{\rm ren} + V_{\rm dim5}$. 
We treat this setup as an effective description, without specifying a particular ultraviolet completion.  
Although other higher-dimensional operators, beyond Eq.~(\ref{eq:Vdim5}), may be present, the most important ones for the mechanism under consideration are those which involve a higher power of the portal operator $(H^\dagger H)^n$, $n > 1$, and which affect the mixing of the lightest new scalar with the standard Higgs boson. 
We focus on the lowest dimensional one,  and assume other higher-dimensional operators to lead to subleading effects on the parameter space associated with the mechanism under study.

The higher-dimensional operator, Eq.~(\ref{eq:Vdim5}), plays a central role.  After symmetry breaking, it contributes differently to the off-diagonal $h$--$s_2$ mass-mixing entry and to the cubic $s_2hh$ interaction. As a result, one can suppress the Higgs admixture of $s_2$, approaching the alignment regime at the price of some level of fine-tuning, while keeping its coupling to a Higgs pair sizable.
 The essential mechanism is therefore to enhance the double-Higgs decays compared to the conventional decays into fermions and gauge bosons. The scalar mass matrix, vacuum conditions, and alignment relations are collected in Appendix~\eqref{app:tad} and~\eqref{app:mass}.

The phenomenologically relevant regime is the approximate alignment limit, where the mixings among $h$ and $s_1$, $s_2$ are small.
This alignment regime is experimentally motivated by Higgs coupling measurements at the LHC, which indicate that the observed 125\,GeV Higgs boson is close to the SM-like.
In this limit, the couplings of $s_1$ and $s_2$ to SM gauge bosons and fermions are suppressed, whereas the scalar cascade interactions can remain unsuppressed. In particular, the trilinear couplings controlling the decays of $s_1$ are
$g_{s_1s_2h} = \kappa v$ (see Eq.~\eqref{gs1s2h}) and $g_{s_1s_2s_2} = 2\lambda_{12}v_1$ (see Eq.~\eqref{gs1s2s2})
so that, depending on kinematics and on the relative sizes of $\kappa$ and $\lambda_{12}v_1$, either $s_1\to hs_2$ or $s_1\to s_2s_2$ can dominate, while all competing SM decay modes remain suppressed in the aligned regime. The VLQ loop-induced decay modes into $gg$ and $\gamma\gamma$ can remain subleading when the scalar cascade couplings, controlled by $\kappa$ and $\lambda_{12}v_1$, are sufficiently large.
\begin{figure*}[t]
  \centering
    \includegraphics[width=0.32\linewidth]{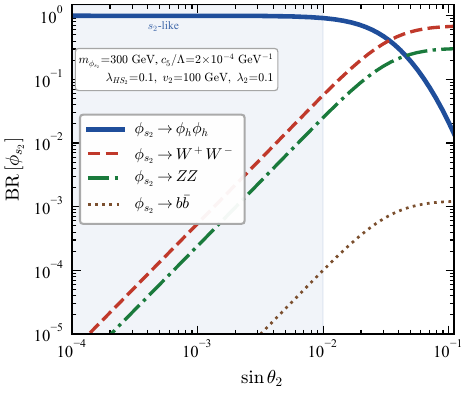} \,\,\,
    \includegraphics[width=0.32\linewidth]{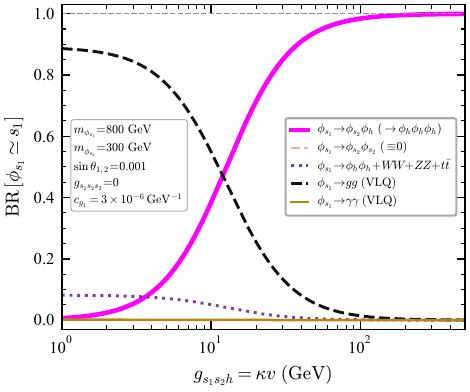}    \,\,\,
    \includegraphics[width=0.32\linewidth]{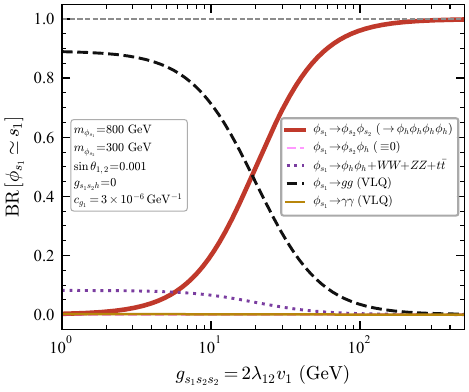}
\caption{
[Left] Branching ratios of  $\phi_{s_2}$ ($m_{\phi_{s_2}} = 300$~GeV) as a function of the $h$--$s_2$ mixing angle $\sin\theta_2$.
In the alignment regime ($\sin\theta_2\lesssim 0.01$, shaded), the mass eigenstates align with the interaction states, and the di-Higgs decay mode dominates over all other channels.
[Middle] Branching ratios of $\phi_{s_1}$ ($m_{\phi_{s_1}} = 800$~GeV)  as a function of the cubic coupling $g_{s_1 s_2 h}$, and [Right] $g_{s_1 s_2 s_2}$. In both panels,  the condition $M^2_{s_1 s_2}=0$ is imposed, and the mixing angles $\sin\theta_1$ and $\sin\theta_2$ are taken to be small, such that $\phi_{s_1}\simeq s_1$, $\phi_{s_2}\simeq s_2$ and $\phi_h \simeq h$. As the cubic couplings increase, the decay channels $s_1\to s_2 h$ (middle) and $s_1\to s_2 s_2$ (right) become dominant, thereby suppressing the loop-induced 
VLQ-mediated $gg$ and $\gamma\gamma$ modes.
The benchmark parameter choices used for illustration are indicated in each panel.
}
  \label{fig:branchingfractions}
\end{figure*}

For the lighter state $s_2$, the interplay between $h$--$s_2$ mass mixing (see Eq.~\eqref{M-hs2}) and cubic $s_2hh$ interaction is particularly important. The relevant quantities are
\begin{align}
M_{hs_2}^2 &= v\!\left(a_2+\kappa v_1+\lambda_{H S_2}v_2\right)
+\frac{c_5}{\Lambda}v^3, 
\label{M-hs2}\\[4pt]
g_{s_2hh}  &= a_2+\kappa v_1 + \lambda_{H S_2}v_2
+3\,\frac{c_5}{\Lambda}v^2 = \frac{M_{hs_2}^2}{v} +2\,\frac{c_5}{\Lambda}v^2 \, .
\label{gs2hh}
\end{align}
Both expressions involve the same portal combination
$(a_2+\kappa v_1+\lambda_{H S_2}v_2)$, but differ in $c_5$ contribution by a factor of three.
This mismatch allows a parametric separation between mixing and decay: one can impose $M_{hs_2}^2\simeq0$ (alignment) while retaining a sizable trilinear coupling,
\begin{equation}
g_{s_2hh}\simeq 2\,\frac{c_5}{\Lambda}v^2.
\end{equation}
As a result, $s_2$ remains highly singlet-like while decaying predominantly into Higgs pairs, with all other two-body decay modes ($ZZ$, $WW$, $f\bar f$) suppressed.
At hadron colliders, $s_2$ can be produced resonantly via the VLQ-induced effective coupling to $gg$, as defined in Eq.~\eqref{s1ggeff_main}. In contrast to typical resonant di-Higgs scenarios, where decays into $WW$, $ZZ$, and $f\bar f$ provide correlated search channels~\cite{DiVita:2017vrr,Cepeda:2019klc, DiMicco:2019ngk,Maltoni:2024dpn, CMS:2024phk,ATLAS:2023vdy,ATLAS:2020fry,CMS:2019kaf}, here these modes are suppressed by alignment, while the VLQ loop-induced $gg$ and $\gamma\gamma$ channels remain subleading for sufficiently large $s_2hh$ coupling. Consequently, $s_2 \to hh$ can dominate with a branching fraction close to 1, making di-Higgs the primary and effectively unique visible probe of $s_2$.

We now focus on triple- and quadruple-Higgs production. A similar mechanism, with a tunable $s_1$ cascade structure, enables a controlled realization of these final states. As this setup involves some degree of parameter tuning, it is expected to be sensitive to renormalization group (RG) effects. We assume the parameters are evaluated at the relevant scale of the theory. Additional quantum corrections may shift the viable parameter space without altering the qualitative physical picture.

To make these features quantitative,
Fig.~\eqref{fig:branchingfractions} shows the branching fractions of
$\phi_{s_2}$ (left) and $\phi_{s_1}$ (center and right) as functions
of $\sin\theta_2$, $g_{s_1s_2h}$, and $g_{s_1s_2s_2}$, respectively, (see Appendix~\eqref{app:widths} for the computational details of the relevant partial decay widths and branching fractions)
for the benchmark
$(m_{\phi_{s_1}},m_{\phi_{s_2}})=(800,300)\,\mathrm{GeV}$, 
$c_5/\Lambda=2\times10^{-4}\,\mathrm{GeV}^{-1}$ and $c_{g_1} = 3 \times 10^{-6}\,\mathrm{GeV}^{-1}$.
For simplicity, in this plot we set the VLQ-induced $s_2 gg$ coupling to zero, as the focus is on triple- and quadruple-Higgs production rather than $s_2$ production. We have verified, however, that even if it is comparable to $c_{g_1}$, the qualitative dominance of the $s_2 \to hh$ decay mode in the low-mixing regime remains unchanged.
In the alignment limit the mass eigenstates align with the interaction states, $\phi_{s_1}\simeq s_1$,
$\phi_{s_2}\simeq s_2$, and $\phi_h\simeq h$.

The left panel shows $\mathrm{BR}(\phi_{s_2})$ as a function of
$\sin\theta_2$.
The coupling $g_{s_2hh}$ and the mixing angle $\theta_2$ are not
independent: both are fixed by the off-diagonal mass entry
$M^2_{hs_2} = \tfrac{1}{2}\sin(2\theta_2)\,(m_{\phi_h}^2 - m_{\phi_{s_2}}^2)$
and $c_5/\Lambda$ through Eq.~\eqref{gs2hh}.
Without loss of generality, we assume that both $\sin2\theta_2$ and $c_5/\Lambda$ are positive.
Since $m_{\phi_h} < m_{\phi_{s_2}}$, one has $M^2_{hs_2} < 0$.
At exact alignment ($\sin\theta_2=0$), $M^2_{hs_2}$ vanishes and
$g_{s_2hh}$ reaches its maximum value $2(c_5/\Lambda)v^2$.
As $\sin\theta_2$ increases, $M^2_{hs_2}$ grows in magnitude and,
since it is negative, acts against the positive $c_5/\Lambda$ term
in $g_{s_2hh}$, progressively reducing the overall coupling and,
consequently, the di-Higgs branching fraction.
Simultaneously, the $WW$, $ZZ$, and $b\bar{b}$ decay modes grow
as $\sin^2\theta_2$.
In the phenomenologically relevant regime $\sin\theta_2\lesssim 0.01$,
one finds $\mathrm{BR}(\phi_{s_2}\to\phi_h\phi_h)\sim 1$.

The center and right panels show $\mathrm{BR}(\phi_{s_1})$ as
functions of $g_{s_1s_2h}=\kappa v$ and
$g_{s_1s_2s_2}=2\lambda_{12}v_1$, respectively,
with $\sin\theta_{1,2}=0.001$ fixed and the alignment condition
$M^2_{s_1s_2}=0$ imposed, corresponding to negligible all pairwise mixings among $h$, $s_1$ and $s_2$. The small mixing region is chosen to suppress $\phi_{s_1} \to ZZ, WW, f \bar{f}$ decay modes. 
At small cubic couplings, the VLQ loop-induced 
$\phi_{s_1}\to gg$ mode dominates; as the couplings grow,
$\phi_{s_1}\to\phi_{s_2}\phi_h$ (center) and
$\phi_{s_1}\to\phi_{s_2}\phi_{s_2}$ (right) rapidly take over,
with branching fractions approaching unity.
Defining the effective branching fractions via the narrow-width
approximation,
$
\mathrm{BR}_{\mathrm{eff}}(hhh)  \equiv
  \mathrm{BR}(\phi_{s_1}\to\phi_{s_2}\phi_h)\,
  \mathrm{BR}(\phi_{s_2}\to\phi_h\phi_h),
$ 
and 
$\mathrm{BR}_{\mathrm{eff}}(hhhh) \equiv
  \mathrm{BR}(\phi_{s_1}\to\phi_{s_2}\phi_{s_2})\,
  \bigl[\mathrm{BR}(\phi_{s_2}\to\phi_h\phi_h)\bigr]^2,$ both can approach  unity in the appropriate regions of parameter space,
establishing triple- and quadruple-Higgs production as the dominant
collider signatures in this scenario.
\subsection{One-real singlet scalar extension}
\label{one-real-singlet}
Resonant multi-Higgs final states can also be realized in a minimal extension of the SM with a single real singlet scalar $S$, supplemented by higher-dimensional operators. 
We consider the effective scalar potential
\begin{align}
\label{VHS}
V(H,S) ={}& -\mu_H^2 H^\dagger H
+ \lambda_H (H^\dagger H)^2
+ \frac{m_1^2}{2} S^2
+ a_1 S H^\dagger H
\nonumber\\
&+ \frac{\lambda_{HS}}{2} S^2 H^\dagger H
+ \frac{\lambda_1}{4} S^4
+ \frac{c_5}{\Lambda} S(H^\dagger H)^2
\nonumber\\
&+ \frac{c_7}{\Lambda^3} S(H^\dagger H)^3 \, ,
\end{align}
with $S=v_s+s$.
If one ignores the dimension seven operator, the same mechanism as before can serve to enhance double Higgs production compared to other two-body standard production channels. Moreover,  
as shown in Appendix~\eqref{sec:single_appendix}, the higher-dimensional operators allow one to approach the alignment limit (no mixing between $s$ and $h$) and suppress the di-Higgs decay $s\to hh$, while retaining sizable couplings to higher-multiplicity Higgs final states. This is achieved through cancellations among operator contributions, which can make $g_{shh} = 0$ without forcing the $s\to hhh$ interaction to vanish. In this limit, the coupling takes the form
$
g_{shhh} = 6 \, c_7 v^3 /\Lambda^3 \, ,
$ (see Eq.~\eqref{gshhh}).
As a result, there exist regions of parameter space in which the $s\to hhh$ decay dominates. 
While additional higher-dimensional operators, such as $\frac{c_9}{\Lambda^5} S(H^\dagger H)^4$, could, in principle, enhance $s\to hhhh$ transitions, achieving a regime in which this channel dominates would require highly non-generic parameter choices and correlated cancellations among multiple higher-dimensional operator contributions.

This illustrates a structural difference between the two scenarios: in the minimal singlet extension, suppressing $s\to hh$ while retaining a sizable $s\to hhh$ coupling requires a cancellation between Wilson coefficients of operators at different mass dimensions (here $c_5$ and $c_7$, see Eq.~\eqref{eq:c5solve}), and dominance of the $s\to hhhh$ channel would require additional correlated cancellations among still higher-dimensional operators. In the two-singlet cascade scenario, by contrast, multi-Higgs dominance is achieved through alignment conditions of the type familiar from extended Higgs sectors  (see, for example, Ref.~\cite{Carena:2015moc}). Both realizations therefore involve some degree of tuning, but the cascade scenario relies on alignment-type relations rather than on cancellations between operators of different dimensions, and it accommodates both $hhh$ and $hhhh$ final states within the same framework. We therefore focus on the cascade realization in the collider phenomenology.
\section{Collider Implications}
\label{sec:pheno}
Having established the regions of parameter space in which multi-Higgs final states dominate, we now turn to the resulting collider signal structure and expected event rates.

\textit{Signal structure and normalization}.—
As established above, the framework admits regions of parameter space in which the effective branching fractions into $hh$, $hhh$, and $hhhh$ final states can approach unity. In this regime, multi-Higgs production constitutes the primary collider signature of the extended scalar sector.
The observable event rates are therefore controlled predominantly by the production cross section of the parent scalar state. At the LHC, the dominant production mechanism is gluon fusion, induced by heavy colored states or, equivalently, by an effective $Xgg$ interaction~\cite{Chen:2017hak,deFlorian:2016spz}. Parametrically, $\sigma(pp \to X) \propto \left(\frac{\alpha_s y_T}{12\pi M_T}\right)^2$, and the signal rate is given by $\sigma_{\text{sig}} = \sigma(pp \to X)\times \text{BR}_{\text{eff}}$. In the region of interest, where $\text{BR}_{\text{eff}} \sim 1$, the multi-Higgs rates are directly set by the production cross section.

As a representative benchmark, we consider a singlet-like scalar with mass $m_{s_1} \simeq 800~\text{GeV}$ and an effective gluon coupling $c_{g_1} \sim 3 \times 10^{-6}~\text{GeV}^{-1}$, yielding a production cross section $\sigma(pp \to X) \sim 15~\text{fb}$ at leading order using \textsc{MadGraph5\_aMC@NLO}~\cite{Alwall:2014hca} at $\sqrt{s} = 14\,\mathrm{TeV}$. For $\text{BR}_{\text{eff}} \sim 1$, this translates into comparable rates for resonant $hh$, $hhh$, and $hhhh$ production, corresponding to potentially observable event yields at the LHC.

While di-Higgs production has been extensively studied both theoretically and experimentally, including dedicated searches at the LHC, higher-multiplicity Higgs final states remain comparatively unexplored. In particular, recent experimental efforts have begun to probe triple-Higgs production, including resonant cascade topologies, while searches for four-Higgs final states are still largely absent. Motivated by this emerging experimental direction, and by the distinctive signal structure of the framework, we focus in the following on the collider phenomenology of $hhh$ and $hhhh$ final states arising from cascade decays.

We begin by outlining the characteristic signal topologies and experimental challenges associated with these high-multiplicity final states. In the cascade realization, the heavy state $s_1$ decays via $s_1 \to h s_2$ or $s_1 \to s_2 s_2$, followed by $s_2 \to hh$, leading to $hhh$ and $hhhh$ final states with a characteristic hierarchical resonance structure.

The phenomenologically dominant signatures arise from $h \to b\bar b$, leading to $6b$ and $8b$ final states with large jet multiplicities and nontrivial combinatorial structure.~\footnote{While we focus on the dominant $h\to b\bar b$ decay mode, subleading channels such as $h\to \gamma\gamma$ and $h\to \tau^+\tau^-$ may provide cleaner but lower-rate signatures, and thus offer complementary search channels. In particular, searches exploiting $h\to \gamma\gamma$ decays have been pursued by the CMS Collaboration in the $4b2\gamma$ final state~\cite{CMS-PAS-HIG-24-015}. For more detail, see the HHH whitepaper~\cite{Abouabid:2024gms}.}
Such final states with high $b$-jet multiplicities are experimentally challenging due to finite $b$-tagging efficiency, which suppresses event-level acceptance, as well as large combinatorial ambiguities and overwhelming QCD multijet backgrounds. These challenges are particularly pronounced in the $8b$ channel. Nevertheless, the presence of correlated invariant-mass structures at multiple scales provides powerful handles that can be exploited for signal discrimination.

The dominant background arises from QCD multijet production, with subleading contributions from $t\bar t+$jets. Although these backgrounds can populate high-multiplicity final states and mimic Higgs candidates combinatorially, they are less likely to reproduce the correlated multi-scale resonance structure of the signal.

\textit{Kinematic structure and topology}.—
Cascade decays of the form $pp \to s_1 \to h s_2$ or $s_1 \to s_2 s_2$, followed by $s_2 \to hh$, give rise to a hierarchical pattern,
\begin{equation}
b\bar{b} \xleftarrow{\;m_{b\bar b}\;} h, \qquad
hh \xleftarrow{\;m_{hh}\;} s_2, \qquad
h^N \xleftarrow{\;m_{h^N}\;} s_1,
\end{equation}
with $N=3,4$, leading to correlated invariant-mass scales in the final state.

A key qualitative distinction arises between asymmetric and symmetric cascade topologies. In the asymmetric configuration $pp \to s_1 \to h s_2 \to hhh$, only one Higgs pair originates from the decay of $s_2$, leading to a broadened intermediate structure due to combinatorial contamination. In contrast, in the symmetric topology $pp \to s_1 \to s_2 s_2 \to hhhh$, the presence of two identical intermediate states introduces correlations that can be exploited to reduce combinatorial ambiguities and improve reconstruction of the intermediate scale.

\textit{Topology-driven reconstruction strategies}.—
The hierarchical structure of the cascade motivates simple reconstruction strategies based on invariant-mass correlations.  For an event with $N$ Higgs bosons, we consider all one-to-one assignments of $b$ and $\bar{b}$ jets and select the configuration minimizing
\begin{equation}
\chi_h^2 = \sum_{i=1}^{N} \frac{\left(m_{b\bar{b}}^{(i)} - m_h\right)^2}{\sigma_h^2},
\end{equation}
where $\sigma_h$ is an analysis-level resolution scale that provides a stable ranking variable for the optimal pairing. This simple $\chi^2$-based procedure serves as a transparent baseline for Higgs reconstruction, in contrast to more sophisticated approaches based on machine-learning techniques used in experimental analyses~\cite{CMS-PAS-HIG-24-012, ATLAS:2024xcs}.

The reconstruction of the intermediate scalar  $s_2$ depends on the topology. In the asymmetric case, combinatorial ambiguity leads to a broadened $m_{hh}$ distribution with a localized signal enhancement. In the symmetric topology, this ambiguity can be reduced by minimizing
\begin{equation}
\Delta_{s_2} = \left| m_{hh}^{(1)} - m_{hh}^{(2)} \right|,
\end{equation}
with the averaged mass
$\bar{m}_{hh} = \frac{m_{hh}^{(1)} + m_{hh}^{(2)}}{2}$
peaking at $m_{s_2}$.

The heavy scalar $s_1$ is reconstructed from the full set of Higgs candidates,
$
m_{h^N}^2 = \left( \sum_{i=1}^{N} p_{h_i} \right)^2.
$
These strategies are intended as illustrative benchmarks of the underlying kinematic structure, which can be further exploited in realistic experimental analyses employing multivariate or
machine-learning techniques.

\textit{Existing constraints and prospects}.—
Existing LHC searches for triple-Higgs production already provide direct constraints on part of the parameter space. The CMS Collaboration has obtained a 95\% CL upper limit of $\sigma(pp\to hhh\to 6b)\lesssim 44~{\rm fb}$~\cite{CMS-PAS-HIG-24-012}. The ATLAS Collaboration has searched for both non-resonant and resonant triple-Higgs production in the $6b$ final state, explicitly targeting cascade topologies, and reports a 95\% CL upper limit of $\sigma(pp\to hhh)\lesssim 59~{\rm fb}$~\cite{ATLAS:2024xcs}.

The $hhhh$ channel is even less constrained experimentally. While $pp\to hhhh\to 8b$ events could in principle contaminate inclusive triple-Higgs searches through partial reconstruction, such effects are expected to be strongly suppressed. Four-Higgs final states therefore remain largely unexplored and constitute a well-motivated target for dedicated searches.

The benchmark rates considered here remain compatible with current constraints, including those from di-Higgs measurements and searches for additional scalar resonances~\cite{ATLAS:2024ish,CMS:2022cpr,DiMicco:2019ngk,Cepeda:2019klc,ATLAS:2023vdy,CMS:2024phk,CMS:2022gjd,CMS:2020tkr,CMS:2026nuu,Collaboration:2928096,CMS:2024pjq,ATLAS:2019qdc,ATLAS:2022xzm,ATLAS:2018uni,ATLAS:2022hwc,ATLAS:2021ifb,ATLAS:2025eii,CMS:2025qit,CMS:2017rpp,CMS-PAS-HIG-24-015,Abouabid:2024gms}.
More generally, the overall signal normalization is controlled by the effective gluon coupling $c_{g_1}$, defined in Eq.~\eqref{cg1}. In the absence of a signal, progressively stronger experimental bounds can be accommodated by moving to smaller values of $c_{g_1}$, thereby reducing the production rate while preserving the characteristic multi-Higgs decay topology.

\textit{Cascade vs.\ direct multi-Higgs production}.—
A qualitatively different behavior arises when multi-Higgs final states originate from direct decays of a single scalar, $s \to hhh$ or $hhhh$, without intermediate resonances. In this case, invariant-mass combinations probe only a single scale, leading to smoother distributions shaped by phase space.  The absence of additional intermediate
mass scale implies that reconstruction strategies based on
correlated di-Higgs structures are not directly applicable,
and dedicated analysis approaches are required.

This distinction provides a direct experimental handle for differentiating between cascade and direct production mechanisms, even when the final states are identical. Additional observables, such as angular separations between $b$-jets, may provide complementary information. A realistic analysis would combine invariant-mass reconstruction, heavy-flavor tagging, and advanced techniques such as boosted Higgs tagging~\cite{Butterworth:2008tr,Larkoski:2017jix} and multivariate methods~\cite{Radovic:2018dip} to enhance sensitivity.

The scenario identified here also highlights an important difference with previous studies of multi-Higgs production in extended scalar sectors. Enhanced triple-Higgs production in such scenarios has been explored, for instance, in the two-real-singlet model with the asymmetric cascade $pp\to h_3\to h_2 h_1\to h_1 h_1 h_1$~\cite{Papaefstathiou:2020lyp}, as well as in simplified narrow-width approaches emphasizing double-resonant factorization~\cite{Papaefstathiou:2025meh}. These studies, which have also motivated recent experimental searches, typically consider regimes where multi-Higgs final states coexist with sizable decays into other SM channels. In contrast, the framework considered here realizes a qualitatively different regime in which $hhh$ and $hhhh$ can become the dominant—and in some cases the only observable signatures over a broad region of parameter space, thereby making resonant multi-Higgs production the primary discovery channel for the underlying scalar dynamics.
\section{Conclusions}
\label{Conclusions}
We have identified a class of scenarios in which multi-Higgs final states, in particular double-, triple- and quadruple-Higgs  channels, can become the dominant collider signatures of extended scalar sectors. Contrary to the conventional expectation that higher-multiplicity Higgs production is strongly suppressed, the decay structure of heavy scalar states can be reorganized such that $hh$, $hhh$ and $hhhh$ production is enhanced, while conventional SM decay modes and lower-multiplicity Higgs channels are parametrically suppressed. In this regime, multi-Higgs final states can provide the primary and, in some cases essentially unique, probes of the extended scalar sector, with potentially sizable production rates at the LHC.

This behavior can be realized through the interplay of alignment and enhanced scalar interactions, including higher-dimensional operators, which suppress competing decay channels while preserving sizable couplings responsible for multi-Higgs production. As a result, both cascade and direct multi-Higgs production can occur with effective branching fractions approaching close to unity.

Identical $hhh$ and $hhhh$ final states can originate from distinct mechanisms with qualitatively different kinematic structures. Cascade topologies generate correlated invariant-mass features associated with intermediate resonances, whereas direct production yields smooth distributions governed by phase space alone, with no intermediate resonant structure imprinted on the final state. These features can be systematically exploited through an event-level hierarchical reconstruction strategy, allowing the underlying decay pattern to be resolved and the origin of the signal to be identified.

Taken together, these results establish  triple- and quadruple-Higgs production as key probes of extended scalar sectors,  beyond the standard di-boson searches. In the regions identified here, where conventional search channels are suppressed, such multi-Higgs signatures can become the primary, and in some cases essentially unique, discovery modes for new scalar dynamics at the LHC and future colliders.

{\bf Acknowledgments.} 
We thank Tim Hobbs, Keisuke Harigaya, Young-Kee Kim, Peiran Li, Zhen Liu, Lian-Tao Wang and the ATLAS group  of the University of Chicago for various insightful discussions.
 SR is supported by the U.S.~Department of Energy under contracts No.\ DEAC02-06CH11357 at the Argonne National Laboratory.  SR would like to thank the University of Chicago, Fermilab and Perimeter Institute where a significant part of this work was carried out. The work of CW\ at the University of Chicago has been supported by the DOE grant DE-SC0013642. 

\noindent {\bf Note added:}
While this work was being completed, Ref.~\cite{Li:2026kqk} appeared, which also explores the role of higher-dimensional Higgs operators in enhancing Higgs-rich final states of heavy resonances. In contrast, we focus on a complementary regime in which conventional decay modes are suppressed, for instance due to alignment, so that multi-Higgs final states can become the dominant collider signatures of the extended scalar sectors.
\appendix
\section{Vacuum structure and tadpole equations}
\label{app:tad}
The full scalar potential is
\begin{equation}
V(H,S_1,S_2)=V_{\rm ren}+\frac{c_5}{\Lambda}S_2(H^\dagger H)^2,
\end{equation}
where $V_{\rm ren}$ is given in Eq.~\eqref{eq:Vren}. Evaluating the potential at the classical background values yields
\begin{align}
V_0 =& -\frac12 \mu_H^2 v^2 + \frac14 \lambda_H v^4
+ \frac12 m_1^2 v_1^2 + \frac12 m_2^2 v_2^2 + \mu_{12}^2 v_1 v_2
\nonumber\\
&+ \frac12 a_1 v_1 v^2 + \frac12 a_2 v_2 v^2 + \frac12 \kappa v_1 v_2 v^2
+ \frac14 \lambda_{H S_1} v_1^2 v^2 
\nonumber\\
&
+ \frac14 \lambda_{H S_2} v_2^2 v^2
+ \frac14 \lambda_1 v_1^4 + \frac14 \lambda_2 v_2^4
+ \frac12 \lambda_{12} v_1^2 v_2^2 + \frac{c_5}{4\Lambda} v_2 v^4.
\end{align}
The vacuum conditions follow from
\begin{equation}
\frac{\partial V_0}{\partial v}=0,\qquad
\frac{\partial V_0}{\partial v_1}=0,\qquad
\frac{\partial V_0}{\partial v_2}=0.
\end{equation}
Explicitly,
\begin{align}
0=&-\mu_H^2+\lambda_H v^2+a_1v_1+a_2v_2+\kappa v_1v_2 
+\frac12\lambda_{H S_1}v_1^2
\nonumber\\
&
+\frac12\lambda_{H S_2}v_2^2+\frac{c_5}{\Lambda}v^2v_2,
\label{eq:tadv}
\\
0=&m_1^2v_1+\mu_{12}^2v_2+\frac12 a_1v^2+\frac12\kappa v^2v_2
+\frac12\lambda_{H S_1}v_1v^2
\nonumber\\
&+
\lambda_1v_1^3+\lambda_{12}v_1v_2^2,
\label{eq:tadv1}
\\
0=&m_2^2v_2+\mu_{12}^2v_1+\frac12 a_2v^2+\frac12\kappa v^2v_1
+\frac12\lambda_{H S_2}v_2v^2
\nonumber\\
&
+\lambda_2v_2^3+\lambda_{12}v_1^2v_2
+\frac{c_5}{4\Lambda}v^4.
\label{eq:tadv2}
\end{align}
A convenient way to solve these equations is to trade the mass parameters $\mu_H^2, m_1^2, m_2^2$ for the vacuum expectation values  and couplings.
\section{Scalar mass matrix, alignment, and cubic couplings}
\label{app:mass}
The full scalar potential is
\begin{equation}
V(H,S_1,S_2)=V_{\rm ren}+\frac{c_5}{\Lambda}S_2(H^\dagger H)^2,
\end{equation}
where $V_{\rm ren}$ is given in Eq.~\eqref{eq:Vren}.
In the $(h,s_1,s_2)$ basis, the $CP$-even mass-squared matrix is
\begin{equation}
\mathcal M^2_{ij}=
\left.\frac{\partial^2 V}{\partial \varphi_i \partial \varphi_j}\right|_{\text{vev}},
\qquad
\varphi_i=(h,s_1,s_2).
\end{equation}
Using the tadpole relations, the entries are
\begin{align}
M_{hh}^2 &= 2\lambda_H v^2 + \frac{2c_5}{\Lambda}v^2v_2,
\\
M_{hs_1}^2 &= v(a_1+\kappa v_2+\lambda_{H S_1}v_1),
\\
\label{M-hs2}
M_{hs_2}^2 &= v(a_2+\kappa v_1+\lambda_{H S_2}v_2)+\frac{c_5}{\Lambda}v^3,
\\
M_{s_1s_1}^2 &= m_1^2+\frac12\lambda_{H S_1}v^2+3\lambda_1v_1^2+\lambda_{12}v_2^2,
\\
M_{s_2s_2}^2 &= m_2^2+\frac12\lambda_{H S_2}v^2+3\lambda_2v_2^2+\lambda_{12}v_1^2,
\\
M_{s_1s_2}^2 &= \mu_{12}^2+\frac12\kappa v^2+2\lambda_{12}v_1v_2.
\end{align}
The physical mass eigenstates are obtained via an orthogonal rotation,
\begin{equation}
\begin{pmatrix}
h\\ s_1\\ s_2
\end{pmatrix}
= R
\begin{pmatrix}
\phi_h\\ \phi_{s_1}\\ \phi_{s_2}
\end{pmatrix},
\label{Rmatrix}
\end{equation}
where $\phi_h$ is identified with the observed 125\,GeV Higgs boson, while $\phi_{s_1}$ and $\phi_{s_2}$ denote the heavier and lighter singlet-like mass eigenstates with masses $m_{s_1}$ and $m_{s_2}$, respectively. In the small-mixing limit, the eigenstates align with the interaction basis, $\phi_h\simeq h$, $\phi_{s_1}\simeq s_1$, and $\phi_{s_2}\simeq s_2$, whereas for larger mixing all three states contain admixtures of $h$, $s_1$, and $s_2$.

In the phenomenologically relevant regime considered here, we impose the condition \(M^2_{s_1 s_2}=0\) for simplicity of the discussion. Although this condition is not necessary for the scenario under consideration, it removes the direct off-diagonal entry between \(s_1\) and \(s_2\) in the interaction basis, thereby simplifying the analysis.
We emphasize, however, that in the full \(3\times 3\) system this
condition alone does not eliminate all \(s_1\)–\(s_2\) mixing: even with
\(M^2_{s_1 s_2} = 0\), the states \(s_1\) and \(s_2\) can mix indirectly
through their common coupling to \(h\) via the entries
\(M^2_{hs_1}\) and \(M^2_{hs_2}\).
In the limit where all off-diagonal entries are small compared to the
relevant diagonal mass splittings, as specified  below, the mixing angles
are perturbatively small and the rotation matrix \(R\) can be expanded
accordingly. To leading order in this small-mixing expansion, the
dominant mixing angles \(\theta_1\) and \(\theta_2\), parametrizing the
\(h\)–\(s_1\) and \(h\)–\(s_2\) mixings respectively, are given by
\begin{align}
    \tan 2\theta_1 & \simeq \frac{2\, M^2_{hs_1}}{M^2_{s_1 s_1} - M^2_{hh}}\,, \label{eq:B10} \\[6pt]
    \tan 2\theta_2 &\simeq \frac{2\, M^2_{hs_2}}{M^2_{s_2 s_2} - M^2_{hh}}\,. \label{eq:B11}
\end{align}
The residual \(s_1\)–\(s_2\) mixing, even with \(M^2_{s_1 s_2} = 0\),
arises at higher order from the two-step path \(s_1 \to h \to s_2\),
and is parametrically of \(\mathcal{O}(\theta_1\theta_2)\). In the
aligned regime this contribution is doubly suppressed and can be
neglected at leading order, justifying the use of \(\theta_1\) and
\(\theta_2\) as the effective parameters controlling the mixing structure~\cite{Carena:2015moc}.

 The aligned regime relevant for our phenomenology is characterized by
\begin{equation}
|M_{hs_1}^2|\ll |M_{s_1s_1}^2-M_{hh}^2|,
\quad
|M_{hs_2}^2|\ll |M_{s_2s_2}^2-M_{hh}^2|,
\end{equation}.
This suggests the approximate alignment conditions
\begin{align}
\label{align-h-s1}
a_1+\kappa v_2+\lambda_{H S_1}v_1 &\simeq 0,
\\
\label{align-h-s2}
a_2+\kappa v_1+\lambda_{H S_2}v_2+\frac{c_5}{\Lambda}v^2 &\simeq 0 \ .
\end{align}
These relations keep the Higgs-like state close to the SM direction while leaving the cubic scalar couplings sufficiently flexible to generate dominant cascade decays.
As emphasized before, for simplicity of the analysis, we also impose the condition 
\begin{equation}
|M_{s_1s_2}^2|\ll |M_{s_1s_1}^2-M_{s_2s_2}^2| \,
\end{equation}
which leads to
\begin{align}
\label{align-s1-s2}
\mu_{12}^2+\frac12\kappa v^2+2\lambda_{12}v_1v_2 &\simeq 0.
\end{align}
The cascade-decay structure in the parameter region of interest is governed by the scalar trilinear interactions
\begin{equation}
s_1 \to h s_2,\qquad s_1 \to s_2 s_2,\qquad s_2 \to hh,
\end{equation}
controlled respectively by the couplings $g_{s_1s_2h}$, $g_{s_1s_2s_2}$, and $g_{s_2hh}$. The relevant interactions are defined through
\begin{align}
\mathcal L_{\rm int} &\supset
-\frac12 g_{s_1hh}s_1h^2
-\frac12 g_{s_2hh}s_2h^2
-g_{s_1s_2h}s_1s_2h
\nonumber\\
&
-\frac12 g_{s_1s_2s_2}s_1s_2^2.
\end{align}
Expanding the potential yields
\begin{align}
g_{s_1hh} &= a_1+\kappa v_2+\lambda_{H S_1}v_1,
\\
\label{eq:gs2hh_full}
g_{s_2hh} &= a_2+\kappa v_1+\lambda_{H S_2}v_2+3\frac{c_5}{\Lambda}v^2,
\\
\label{gs1s2h}
g_{s_1s_2h} &= \kappa v,
\\
\label{gs1s2s2}
g_{s_1s_2s_2} &= 2\lambda_{12}v_1.
\end{align}
In the alignment limit, using Eq.~\eqref{align-h-s1}, one finds that 
\begin{equation}
g_{s_1hh}\simeq 0.
\end{equation}
The important structural point is that the combination controlling $g_{s_2hh}$ is not identical to the one controlling the $h$--$s_2$ mass mixing in Eq.~\eqref{M-hs2}. It is therefore possible to suppress the Higgs admixture of $s_2$ while keeping the cubic coupling to a Higgs pair sizable. Using the alignment relation in Eq.~(\ref{align-h-s2}), one obtains
\begin{equation}
\label{eq:gs2hh_align}
g_{s_2hh}\simeq 2\frac{c_5}{\Lambda}v^2.
\end{equation}
This is the central mechanism underlying the dominance of $s_2\to hh$ in our setup.
\section{One-singlet extension scenario for dominant multi-Higgs decays}
\label{sec:single_appendix}
The phenomenologically motivated potential is expressed in Eq.~\eqref{VHS}. 
The $CP$-even mass-squared matrix in the $(h,s)$ basis contains the off-diagonal entry
\begin{equation}
M_{hs}^2
=
v\left(
a_1+\lambda_{HS}v_s+\frac{c_5}{\Lambda}v^2
+\frac{3c_7}{4\Lambda^3}v^4
\right).
\end{equation}
Exact alignment is obtained by imposing $M_{hs}^2=0$, which fixes
\begin{equation}
a_1
=
-\lambda_{HS}v_s
-\frac{c_5}{\Lambda}v^2
-\frac{3c_7}{4\Lambda^3}v^4 \, .
\label{eq:single_align}
\end{equation}
The interactions relevant for resonant multi-Higgs decays are
\begin{align}
g_{shh}
&=
a_1+\lambda_{HS}v_s+\frac{3c_5}{\Lambda}v^2
+\frac{15c_7}{4\Lambda^3}v^4 \, , \\
g_{shhh}
&=
\frac{3}{\Lambda^3}
\left(
5c_7 v^3+2c_5 v\Lambda^2
\right).
\end{align}
After imposing the alignment condition in Eq.~(\ref{eq:single_align}), these become
\begin{align}
g_{shh}^{\rm align}
&=
\frac{1}{\Lambda^3}
\left(
3c_7 v^4+2c_5 v^2\Lambda^2
\right), \\
g_{shhh}^{\rm align}
&=
\frac{3}{\Lambda^3}
\left(
5c_7 v^3+2c_5 v\Lambda^2
\right).
\end{align}
This shows explicitly that alignment alone does not suppress the di-Higgs channel: even for vanishing $h$--$s$ mixing, $g_{shh}$ remains nonzero in general.
Now, suppressing the di-Higgs channel therefore requires an additional condition beyond alignment,
\[
g_{shh}^{\rm align}=0.
\]
Solving this for $c_5$ gives
\begin{equation}
c_5
=
-\frac{3 c_7 v^2}{2\Lambda^2} \, .
\label{eq:c5solve}
\end{equation}
After this substitution, $g_{shhh}$ reduces to
\begin{equation}
\label{gshhh}
g_{shhh}
=
\frac{6 c_7 v^3}{\Lambda^3} \, .
\end{equation}
Therefore, the cancellation of the di-Higgs coupling does not force the $s\to hhh$ interaction to vanish.

In principle, extending the EFT with higher-dimensional operators such as 
$\frac{c_9}{\Lambda^5} S(H^\dagger H)^4$ can generate $s\to hhhh$ interactions. 
However, realizing a regime in which both $g_{shh}$ and $g_{shhh}$ are suppressed 
while $g_{shhhh}$ remains sizable (for example, one finds $g_{shhhh} = 24 \, c_9 \, v^4/\Lambda^5$ in this limit) requires highly correlated relations among 
multiple operator coefficients, and is therefore non-generic.
For this reason, while the EFT formally allows the construction of higher-multiplicity Higgs final states in the minimal singlet scenario, their realization relies on nontrivial cancellations. In contrast, the two-singlet cascade scenario discussed in the main text achieves multi-Higgs dominance in a relatively more natural and robust manner.
\section{Partial decay widths}
\label{app:widths}
This appendix collects the partial-width formulas used to produce
Fig.~\eqref{fig:branchingfractions}.
The two-body phase-space momentum for a decay $M\to m_1+m_2$ is
\begin{equation}
p_{\rm cm}(M;\,m_1,m_2)
= \frac{\sqrt{\lambda(M^2,\,m_1^2,\,m_2^2)}}{2M}\,,
\label{eq:pcm}
\end{equation}
where $\lambda(a,b,c)=a^2+b^2+c^2-2ab-2bc-2ca$ is the
K\"{a}ll\'{e}n function, with the threshold condition
$M\geq m_1+m_2$.

The di-Higgs decay rate of $\phi_{s_2}$ depends on the physical 
$\phi_{s_2}\phi_h\phi_h$ coupling, $G_{\rm eff}$, which is obtained 
by expressing the interaction-basis fields in terms of the mass 
eigenstates via the rotation $R$ of Eq.~\eqref{Rmatrix} and extracting 
the coefficient of the $\phi_{s_2}\phi_h^2$ operator. In the limit 
$\theta_1, \theta_{s_1s_2} \to 0$ , the $\phi_{s_2}\phi_h\phi_h$ vertex receives contributions only 
from the $h$--$s_2$ mixing sector, and the rotation $R$ effectively 
reduces to a $2\times 2$ rotation in the $(h, s_2)$ subspace parametrized 
by $\theta_2$ alone. 
Including all four cubic interactions of the scalar potential,
one finds
\begin{align}
G_{\rm eff}
&= g_{s_2hh}\,c_{\theta_2}(1-3s_{\theta_2}^2)
+ g_{hhh}\,s_{\theta_2}c_{\theta_2}^2 \nonumber \\&
+ g_{hs_2s_2}\,s_{\theta_2}(s_{\theta_2}^2-2c_{\theta_2}^2)
+ g_{s_2s_2s_2}\,s_{\theta_2}^2 c_{\theta_2}
\label{eq:Geff}
\end{align}
where $c_{\theta_2}\equiv\cos\theta_2$,
$s_{\theta_2}\equiv\sin\theta_2$.
The partial width is
\begin{equation}
\Gamma(\phi_{s_2}\to\phi_h\phi_h)
= \frac{G_{\rm eff}^2\;
        p_{\rm cm}(m_{\phi_{s_2}};\,m_{\phi_h},m_{\phi_h})}
       {16\pi\,m_{\phi_{s_2}}^2}\, .
\label{eq:Ghh}
\end{equation}
In the alignment limit ($\sin\theta_2\to 0$),
$G_{\rm eff}\to g_{s_2hh}\simeq 2(c_5/\Lambda)v^2$,
as found in Eq.~\eqref{eq:gs2hh_align}.

The decay modes of $\phi_{s_2}$ into various SM particles are controlled by its
doublet admixture, $\sin\theta_2$, and are therefore universally suppressed
by $\sin^2\theta_2$. The corresponding partial widths are given by
\begin{align}
\Gamma(\phi_{s_2}\to W^+W^-)
&= \frac{\sin^2\!\theta_2\;G_F\,m_{\phi_{s_2}}^3}{8\pi\sqrt{2}}
   \left(1-4x_W+12x_W^2\right) \nonumber \\& \times \sqrt{1-4x_W}\,,~~~~~~~~~~~~~~~~~~~~~~~~~~~~~~~~~~~~~~~~~~~~~~
\label{eq:GWW}\\
\Gamma(\phi_{s_2}\to ZZ)
&= \frac{\sin^2\!\theta_2\;G_F\,m_{\phi_{s_2}}^3}{16\pi\sqrt{2}}
   \left(1-4x_Z+12x_Z^2\right) \nonumber \\& \times \sqrt{1-4x_Z}\,,~~~~~~~~~~~~~~~~~~~~~~~~~~~~~~~~~~~~~~~~~~~~~~
\label{eq:GZZ}\\
\Gamma(\phi_{s_2}\to f\bar{f})
&= 
   \frac{N_c\,\sin^2\!\theta_2 (m_f/v)^2\,m_{\phi_{s_2}}}{8\pi}
   \left(1-\frac{4m_f^2}{m_{\phi_{s_2}}^2}\right)^{\!\!3/2} ,~~~~~~~~~~~~~~~~~~~~~~~~~~~~~~~~~~~~~~~~~~~~~~\nonumber \\ 
\label{eq:Gff}
\end{align}
where $x_{W,Z}\equiv(m_{W,Z}/m_{\phi_{s_2}})^2$ and
$N_c=3\,(1)$ for quarks (leptons).
In the alignment limit $\sin\theta_2\to 0$, the direct coupling of
$\phi_{s_2}$ to SM gauge bosons and fermions vanishes, while the
trilinear coupling $\phi_{s_2}\phi_h\phi_h$ can remain nonzero.
Consequently, in the framework considered here, the decay modes into SM
gauge bosons and fermions are suppressed by alignment, while
$\mathrm{BR}(\phi_{s_2}\to\phi_h\phi_h)\to 1$.

In the alignment limit, $\phi_{s_1}\to\phi_{s_2}\phi_h$ and $\phi_{s_1}\to\phi_{s_2}\phi_{s_2}$ decay processes are mostly controlled by the 
cubic terms $g_{s_1s_2h} = \kappa v$ and $g_{s_1s_2s_2} = 2\lambda_{12}v_1$,
corrections from the full rotation matrix $R$ enter at 
$\mathcal{O}(\sin^2\theta_{1,2})$ and are negligible for 
$\sin\theta_{1,2} \lesssim 0.001$.
In this limit, the partial widths for the two cascade channels are
\begin{align}
\Gamma(\phi_{s_1}\to\phi_{s_2}\phi_h)
&\simeq \frac{g_{s_1s_2h}^2\;
   p_{\rm cm}(m_{\phi_{s_1}};\,m_{\phi_{s_2}},m_{\phi_h})}
   {8\pi\,m_{\phi_{s_1}}^2},
\label{eq:Ghs2}\\[6pt]
\Gamma(\phi_{s_1}\to\phi_{s_2}\phi_{s_2})
&\simeq \frac{g_{s_1s_2s_2}^2\;
   p_{\rm cm}(m_{\phi_{s_1}};\,m_{\phi_{s_2}},m_{\phi_{s_2}})}
   {16\pi\,m_{\phi_{s_1}}^2}  \, .
\label{eq:Gs2s2}
\end{align}
The decay modes of $\phi_{s_1}$ into $WW$, $ZZ$, and $f\bar{f}$ are analogous to those in Eqs.~\eqref{eq:GWW},~\eqref{eq:GZZ}, and~\eqref{eq:Gff}, with the replacements $\sin\theta_2 \to \sin\theta_1$ and $m_{\phi_{s_2}} \to m_{\phi_{s_1}}$.
The VLQ loop-induced widths $\Gamma(\phi_{s_1}\to gg)$ and
$\Gamma(\phi_{s_1}\to\gamma\gamma)$ are given in
Eqs.~\eqref{eq:Gams1gg} and~\eqref{eq:Gams1AA}.
\section{VLQ-induced gluon coupling}
\label{app:gg}
The VLQ induces the effective interaction
\begin{equation}
\label{s1ggeff}
\mathcal L_{\rm eff}\supset c_{g1}\, s_1\, G^a_{\mu\nu}G^{a\mu\nu},
\end{equation}
with
\begin{equation}
\label{cg1}
c_{g1}=
\frac{\alpha_s}{16\pi}\frac{y_{T,1}}{M_T}A_{1/2}(\tau_T),
\qquad
\tau_T=\frac{4M_T^2}{m_{s_1}^2}.
\end{equation}
The standard spin-$1/2$ loop function is
\begin{equation}
A_{1/2}(\tau)=2\tau\big[1+(1-\tau)f(\tau)\big],
\end{equation}
where
\begin{equation}
f(\tau)=
\begin{cases}
\arcsin^2(\tau^{-1/2}), & \tau\ge 1,\\[4pt]
-\dfrac14\left[\ln\!\left(\dfrac{1+\sqrt{1-\tau}}{1-\sqrt{1-\tau}}\right)-i\pi\right]^2,
& \tau<1.
\end{cases}
\end{equation}
In the heavy-fermion limit,
\begin{equation}
A_{1/2}(\tau)\to \frac43 \, ,
\end{equation}
 and combining Eqs.~\eqref{s1ggeff} and~\eqref{cg1}, one recovers the form given in Eq.~\eqref{s1ggeff_main}.
Thus, the resonant production cross section scales parametrically as
\begin{equation}
\sigma(pp\to s_1)\propto
\left|
\frac{y_{T,1}}{M_T}A_{1/2}(\tau_T)
\right|^2 \, .
\end{equation}
Current LHC searches push top-like vector-like quarks into the TeV range, with pair-production limits typically excluding masses around $1.4$--$1.6$ TeV and single-production searches constraining the corresponding electroweak mixing parameters for masses up to about $2$ TeV, depending on the assumed couplings and branching fractions~\cite{CMS:2024qdd, ATLAS:2024mrr, Benbrik:2024fku, OropezaBarrera:2024ddv}. In the present work we therefore treat $c_{g1}$ as an effective low-energy parameter generated by integrating out a heavy VLQ. For the benchmark value $c_{g1} \sim 3 \times 10^{-6}\,\mathrm{GeV}^{-1}$, the heavy-fermion expression in Eq.~(\ref{cg1}) implies $M_T$ in the TeV range for the Yukawa coupling $y_{T,1} \sim 1$. This remains compatible with a perturbative effective description and should be viewed as an illustrative benchmark rather than a complete UV fit to current direct-search limits.

The VLQ effective operators induce decays which are given by,
\begin{align}
\Gamma(s_1\to gg)          &= \frac{2\,c_{g_1}^2\,m_{s_1}^3}{\pi},
  \label{eq:Gams1gg}\\[3pt]
\Gamma(s_1\to\gamma\gamma)  &= \frac{c_\gamma^2\,m_{s_1}^3}{4\pi},
\qquad
c_\gamma = 2\frac{\alpha}{\alpha_s}Q_T^2 N_c\,c_{g_1}.
  \label{eq:Gams1AA}
\end{align}
The corresponding expressions for the $s_2$ interactions follow from the replacements $m_{s_1} \to m_{s_2}$ and $y_{T,1} \to y_{T,2}$ in the above relations.

\bibliography{paper.bib}

@article{ATLAS:2022xzm,
    author = "Aad, Georges and others",
    collaboration = "ATLAS",
    title = "{Search for resonant and non-resonant Higgs boson pair production in the $ b\overline{b}{\tau}^{+}{\tau}^{-} $ decay channel using 13 TeV pp collision data from the ATLAS detector}",
    eprint = "2209.10910",
    archivePrefix = "arXiv",
    primaryClass = "hep-ex",
    reportNumber = "CERN-EP-2022-109",
    doi = "10.1007/JHEP07(2023)040",
    journal = "JHEP",
    volume = "07",
    pages = "040",
    year = "2023"
}

@article{ATLAS:2018uni,
    author = "Aaboud, Morad and others",
    collaboration = "ATLAS",
    title = "{Search for resonant and non-resonant Higgs boson pair production in the ${b\bar{b}\tau^+\tau^-}$ decay channel in $pp$ collisions at $\sqrt{s}=13$ TeV with the ATLAS detector}",
    eprint = "1808.00336",
    archivePrefix = "arXiv",
    primaryClass = "hep-ex",
    reportNumber = "CERN-EP-2018-164",
    doi = "10.1103/PhysRevLett.121.191801",
    journal = "Phys. Rev. Lett.",
    volume = "121",
    number = "19",
    pages = "191801",
    year = "2018",
    note = "[Erratum: Phys.Rev.Lett. 122, 089901 (2019)]"
}

@article{CMS:2017rpp,
    author = "Sirunyan, Albert M and others",
    collaboration = "CMS",
    title = "{Search for resonant and nonresonant Higgs boson pair production in the $ \mathrm{b}\overline{\mathrm{b}}\mathit{\ell \nu \ell \nu } $ final state in proton-proton collisions at $ \sqrt{s}=13 $ TeV}",
    eprint = "1708.04188",
    archivePrefix = "arXiv",
    primaryClass = "hep-ex",
    reportNumber = "CMS-HIG-17-006, CERN-EP-2017-168",
    doi = "10.1007/JHEP01(2018)054",
    journal = "JHEP",
    volume = "01",
    pages = "054",
    year = "2018"
}

@article{ATLAS:2012yve,
    author = "Aad, Georges and others",
    collaboration = "ATLAS",
    title = "{Observation of a new particle in the search for the Standard Model Higgs boson with the ATLAS detector at the LHC}",
    eprint = "1207.7214",
    archivePrefix = "arXiv",
    primaryClass = "hep-ex",
    reportNumber = "CERN-PH-EP-2012-218",
    doi = "10.1016/j.physletb.2012.08.020",
    journal = "Phys. Lett. B",
    volume = "716",
    pages = "1--29",
    year = "2012"
}

@article{CMS:2012qbp,
    author = "Chatrchyan, Serguei and others",
    collaboration = "CMS",
    title = "{Observation of a New Boson at a Mass of 125 GeV with the CMS Experiment at the LHC}",
    eprint = "1207.7235",
    archivePrefix = "arXiv",
    primaryClass = "hep-ex",
    reportNumber = "CMS-HIG-12-028, CERN-PH-EP-2012-220",
    doi = "10.1016/j.physletb.2012.08.021",
    journal = "Phys. Lett. B",
    volume = "716",
    pages = "30--61",
    year = "2012"
}

@article{Cepeda:2019klc,
    author = "Cepeda, M. and others",
    editor = "Dainese, Andrea and Mangano, Michelangelo and Meyer, Andreas B. and Nisati, Aleandro and Salam, Gavin and Vesterinen, Mika Anton",
    title = "{Report from Working Group 2}: {Higgs Physics at the HL-LHC and HE-LHC}",
    eprint = "1902.00134",
    archivePrefix = "arXiv",
    primaryClass = "hep-ph",
    reportNumber = "CERN-LPCC-2018-04",
    doi = "10.23731/CYRM-2019-007.221",
    journal = "CERN Yellow Rep. Monogr.",
    volume = "7",
    pages = "221--584",
    year = "2019"
}

@article{deBlas:2022aow,
    author = "de Blas, Jorge and Gu, Jiayin and Liu, Zhen",
    title = "{Higgs boson precision measurements at a 125~GeV muon collider}",
    eprint = "2203.04324",
    archivePrefix = "arXiv",
    primaryClass = "hep-ph",
    doi = "10.1103/PhysRevD.106.073007",
    journal = "Phys. Rev. D",
    volume = "106",
    number = "7",
    pages = "073007",
    year = "2022"
}

@article{Forslund:2023reu,
    author = "Forslund, Matthew and Meade, Patrick",
    title = "{Precision Higgs width and couplings with a high energy muon collider}",
    eprint = "2308.02633",
    archivePrefix = "arXiv",
    primaryClass = "hep-ph",
    doi = "10.1007/JHEP01(2024)182",
    journal = "JHEP",
    volume = "01",
    pages = "182",
    year = "2024"
}

@article{ATLAS:2024ish, 
    author = "Aad, Georges and others",
    collaboration = "ATLAS",
    title = "{Combination of Searches for Higgs Boson Pair Production in pp Collisions at s=13{\,}{\,}TeV with the ATLAS Detector}",
    eprint = "2406.09971",
    archivePrefix = "arXiv",
    primaryClass = "hep-ex",
    reportNumber = "CERN-EP-2024-160",
    doi = "10.1103/PhysRevLett.133.101801",
    journal = "Phys. Rev. Lett.",
    volume = "133",
    number = "10",
    pages = "101801",
    year = "2024"
}

@article{CMS:2022cpr,
    author = "Tumasyan, Armen and others",
    collaboration = "CMS",
    title = "{Search for Higgs Boson Pair Production in the Four b Quark Final State in Proton-Proton Collisions at s=13{\,}{\,}TeV}",
    eprint = "2202.09617",
    archivePrefix = "arXiv",
    primaryClass = "hep-ex",
    reportNumber = "CMS-HIG-20-005, CERN-EP-2022-004",
    doi = "10.1103/PhysRevLett.129.081802",
    journal = "Phys. Rev. Lett.",
    volume = "129",
    number = "8",
    pages = "081802",
    year = "2022"
}

@article{Alwall:2014hca,
    author = "Alwall, J. and Frederix, R. and Frixione, S. and Hirschi, V. and Maltoni, F. and Mattelaer, O. and Shao, H. -S. and Stelzer, T. and Torrielli, P. and Zaro, M.",
    title = "{The automated computation of tree-level and next-to-leading order differential cross sections, and their matching to parton shower simulations}",
    eprint = "1405.0301",
    archivePrefix = "arXiv",
    primaryClass = "hep-ph",
    reportNumber = "CERN-PH-TH-2014-064, CP3-14-18, LPN14-066, MCNET-14-09, ZU-TH-14-14",
    doi = "10.1007/JHEP07(2014)079",
    journal = "JHEP",
    volume = "07",
    pages = "079",
    year = "2014"
}

@article{Papaefstathiou:2025meh, 
    author = "Papaefstathiou, Andreas and Tetlalmatzi-Xolocotzi, Gilberto",
    title = "{Triple Higgs Boson Production with Two Heavy Scalars at the LHC via a Simplified Approach}",
    eprint = "2501.14866",
    archivePrefix = "arXiv",
    primaryClass = "hep-ph",
    reportNumber = "SI-HEP-2024-34, P3H-25-002",
    month = "1",
    year = "2025"
}

@article{DiMicco:2019ngk,
    author = "Alison, J. and others",
    editor = "Di Micco, Biagio and Gouzevitch, Maxime and Mazzitelli, Javier and Vernieri, Caterina",
    title = "{Higgs boson potential at colliders: Status and perspectives}",
    eprint = "1910.00012",
    archivePrefix = "arXiv",
    primaryClass = "hep-ph",
    reportNumber = "FERMILAB-CONF-19-468-E-T, LHCXSWG-2019-005",
    doi = "10.1016/j.revip.2020.100045",
    journal = "Rev. Phys.",
    volume = "5",
    pages = "100045",
    year = "2020"
}

@article{ATLAS:2024xcs,
    author = "Aad, Georges and others",
    collaboration = "ATLAS",
    title = "{Search for triple Higgs boson production in the 6b final state using pp collisions at s=13{\,}{\,}TeV with the ATLAS detector}",
    eprint = "2411.02040",
    archivePrefix = "arXiv",
    primaryClass = "hep-ex",
    reportNumber = "CERN-EP-2024-285",
    doi = "10.1103/PhysRevD.111.032006",
    journal = "Phys. Rev. D",
    volume = "111",
    number = "3",
    pages = "032006",
    year = "2025"
}

@article{ATLAS:2023vdy,
    author = "Aad, Georges and others",
    collaboration = "ATLAS",
    title = "{Combination of Searches for Resonant Higgs Boson Pair Production Using pp Collisions at s=13{\,}{\,}TeV with the ATLAS Detector}",
    eprint = "2311.15956",
    archivePrefix = "arXiv",
    primaryClass = "hep-ex",
    reportNumber = "CERN-EP-2023-271",
    doi = "10.1103/PhysRevLett.132.231801",
    journal = "Phys. Rev. Lett.",
    volume = "132",
    number = "23",
    pages = "231801",
    year = "2024"
}

@article{CMS:2024phk,
    author = "Hayrapetyan, Aram and others",
    collaboration = "CMS",
    title = "{Searches for Higgs boson production through decays of heavy resonances}",
    eprint = "2403.16926",
    archivePrefix = "arXiv",
    primaryClass = "hep-ex",
    reportNumber = "CMS-B2G-23-002, CERN-EP-2024-062",
    doi = "10.1016/j.physrep.2024.09.004",
    journal = "Phys. Rept.",
    volume = "1115",
    pages = "368--447",
    year = "2025"
}

@article{CMS:2024pjq,
    author = "Tumasyan, Armen and others",
    collaboration = "CMS",
    title = "{Search for resonant pair production of Higgs bosons in the $ \textrm{b}\overline{\textrm{b}}\textrm{b}\overline{\textrm{b}} $ final state using large-area jets in proton-proton collisions at $ \sqrt{s} $ = 13 TeV}",
    eprint = "2407.13872",
    archivePrefix = "arXiv",
    primaryClass = "hep-ex",
    reportNumber = "CMS-B2G-20-004, CERN-EP-2024-030",
    doi = "10.1007/JHEP02(2025)040",
    journal = "JHEP",
    volume = "02",
    pages = "040",
    year = "2025"
}

@article{CMS:2026nuu,
    author = "Aad, Georges and others",
    collaboration = "{CMS, ATLAS}",
    title = "{Combination of ATLAS and CMS searches for Higgs boson pair production at $\sqrt{s} = 13$ TeV}",
    eprint = "2602.23991",
    archivePrefix = "arXiv",
    primaryClass = "hep-ex",
    reportNumber = "CERN-EP-2026-011",
    month = "2",
    year = "2026"
}

@article{AlAli:2021let,
    author = "Al Ali, Hind and others",
    title = "{The muon Smasher{\textquoteright}s guide}",
    eprint = "2103.14043",
    archivePrefix = "arXiv",
    primaryClass = "hep-ph",
    doi = "10.1088/1361-6633/ac6678",
    journal = "Rept. Prog. Phys.",
    volume = "85",
    number = "8",
    pages = "084201",
    year = "2022"
}

@article{DiVita:2017vrr,
    author = "Di Vita, Stefano and Durieux, Gauthier and Grojean, Christophe and Gu, Jiayin and Liu, Zhen and Panico, Giuliano and Riembau, Marc and Vantalon, Thibaud",
    title = "{A global view on the Higgs self-coupling at lepton colliders}",
    eprint = "1711.03978",
    archivePrefix = "arXiv",
    primaryClass = "hep-ph",
    reportNumber = "DESY-17-131, FERMILAB-PUB-17-462-T",
    doi = "10.1007/JHEP02(2018)178",
    journal = "JHEP",
    volume = "02",
    pages = "178",
    year = "2018"
}

@article{ATLAS:2021ifb,
    author = "Aad, Georges and others",
    collaboration = "ATLAS",
    title = "{Search for Higgs boson pair production in the two bottom quarks plus two photons final state in $pp$ collisions at $\sqrt{s}=13$ TeV with the ATLAS detector}",
    eprint = "2112.11876",
    archivePrefix = "arXiv",
    primaryClass = "hep-ex",
    reportNumber = "CERN-EP-2021-180",
    doi = "10.1103/PhysRevD.106.052001",
    journal = "Phys. Rev. D",
    volume = "106",
    number = "5",
    pages = "052001",
    year = "2022"
}

@article{Tian:2013yda,
    author = "Tian, Junping and Fujii, Keisuke",
    collaboration = "ILD",
    title = "{Measurement of Higgs couplings and self-coupling at the ILC}",
    eprint = "1311.6528",
    archivePrefix = "arXiv",
    primaryClass = "hep-ph",
    doi = "10.22323/1.180.0316",
    journal = "PoS",
    volume = "EPS-HEP2013",
    pages = "316",
    year = "2013"
}

@article{Barklow:2017suo,
    author = "Barklow, Tim and Fujii, Keisuke and Jung, Sunghoon and Karl, Robert and List, Jenny and Ogawa, Tomohisa and Peskin, Michael E. and Tian, Junping",
    title = "{Improved Formalism for Precision Higgs Coupling Fits}",
    eprint = "1708.08912",
    archivePrefix = "arXiv",
    primaryClass = "hep-ph",
    reportNumber = "DESY-17-120, KEK-PREPRINT-2017-22, SLAC-PUB-17129",
    doi = "10.1103/PhysRevD.97.053003",
    journal = "Phys. Rev. D",
    volume = "97",
    number = "5",
    pages = "053003",
    year = "2018"
}

@article{dEnterria:2017dac,
    author = "d'Enterria, David",
    title = "{Higgs physics at the Future Circular Collider}",
    eprint = "1701.02663",
    archivePrefix = "arXiv",
    primaryClass = "hep-ex",
    doi = "10.22323/1.282.0434",
    journal = "PoS",
    volume = "ICHEP2016",
    pages = "434",
    year = "2017"
}

@article{Forslund:2022xjq,
    author = "Forslund, Matthew and Meade, Patrick",
    title = "{High precision higgs from high energy muon colliders}",
    eprint = "2203.09425",
    archivePrefix = "arXiv",
    primaryClass = "hep-ph",
    reportNumber = "YITP-SB-22-11",
    doi = "10.1007/JHEP08(2022)185",
    journal = "JHEP",
    volume = "08",
    pages = "185",
    year = "2022"
}

@article{Li:2024joa,
    author = "Li, Peiran and Liu, Zhen and Lyu, Kun-Feng",
    title = "{Higgs boson width and couplings at high energy muon colliders with forward muon detection}",
    eprint = "2401.08756",
    archivePrefix = "arXiv",
    primaryClass = "hep-ph",
    doi = "10.1103/PhysRevD.109.073009",
    journal = "Phys. Rev. D",
    volume = "109",
    number = "7",
    pages = "073009",
    year = "2024"
}

@article{Maltoni:2024dpn,
    author = "Maltoni, Fabio and Ventura, Giuseppe and Vryonidou, Eleni",
    title = "{Impact of SMEFT renormalisation group running on Higgs production at the LHC}",
    eprint = "2406.06670",
    archivePrefix = "arXiv",
    primaryClass = "hep-ph",
    doi = "10.1007/JHEP12(2024)183",
    journal = "JHEP",
    volume = "12",
    pages = "183",
    year = "2024"
}

@techreport{CMS-PAS-HIG-24-012,
  author      = "{CMS Collaboration}",
  title       = "{Search for nonresonant triple Higgs boson production in the six b-quark final state in proton-proton collisions at 13 TeV}",
  institution = "CERN",
  number      = "CMS-PAS-HIG-24-012",
  year        = "2025",
  url         = "https://cds.cern.ch/record/2945361"
}

@article{ATLAS:2020fry,
    author = "Aad, Georges and others",
    collaboration = "ATLAS",
    title = "{Search for heavy diboson resonances in semileptonic final states in pp collisions at $\sqrt{s}=13$ TeV with the ATLAS detector}",
    eprint = "2004.14636",
    archivePrefix = "arXiv",
    primaryClass = "hep-ex",
    reportNumber = "CERN-EP-2020-049",
    doi = "10.1140/epjc/s10052-020-08554-y",
    journal = "Eur. Phys. J. C",
    volume = "80",
    number = "12",
    pages = "1165",
    year = "2020"
}

@article{CMS:2019kaf,
    author = "Sirunyan, Albert M and others",
    collaboration = "CMS",
    title = "{Combination of CMS searches for heavy resonances decaying to pairs of bosons or leptons}",
    eprint = "1906.00057",
    archivePrefix = "arXiv",
    primaryClass = "hep-ex",
    reportNumber = "CMS-B2G-18-006, CERN-EP-2019-110",
    doi = "10.1016/j.physletb.2019.134952",
    journal = "Phys. Lett. B",
    volume = "798",
    pages = "134952",
    year = "2019"
}

@article{Chen:2017hak,
    author = "Chen, Chien-Yi and Dawson, S. and Furlan, Elisabetta",
    title = "{Vectorlike fermions and Higgs effective field theory revisited}",
    eprint = "1703.06134",
    archivePrefix = "arXiv",
    primaryClass = "hep-ph",
    doi = "10.1103/PhysRevD.96.015006",
    journal = "Phys. Rev. D",
    volume = "96",
    number = "1",
    pages = "015006",
    year = "2017"
}

@article{Cornwall:1974km,
    author = "Cornwall, John M. and Levin, David N. and Tiktopoulos, George",
    title = "{Derivation of Gauge Invariance from High-Energy Unitarity Bounds on the s Matrix}",
    reportNumber = "UCLA-74-TEP-2",
    doi = "10.1103/PhysRevD.10.1145",
    journal = "Phys. Rev. D",
    volume = "10",
    pages = "1145",
    year = "1974",
    note = "[Erratum: Phys.Rev.D 11, 972 (1975)]"
}

@article{Lee:1977eg,
    author = "Lee, Benjamin W. and Quigg, C. and Thacker, H. B.",
    title = "{Weak Interactions at Very High-Energies: The Role of the Higgs Boson Mass}",
    reportNumber = "FERMILAB-PUB-77-030-T",
    doi = "10.1103/PhysRevD.16.1519",
    journal = "Phys. Rev. D",
    volume = "16",
    pages = "1519",
    year = "1977"
}

@article{Vayonakis:1976vz, 
    author = "Vayonakis, C. E.",
    title = "{Born Helicity Amplitudes and Cross-Sections in Nonabelian Gauge Theories}",
    reportNumber = "Print-76-0770 (SUSSEX)",
    doi = "10.1007/BF02746538",
    journal = "Lett. Nuovo Cim.",
    volume = "17",
    pages = "383",
    year = "1976"
}

@article{ATLAS:2022hwc,
    author = "Aad, Georges and others",
    collaboration = "ATLAS",
    title = "{Search for resonant pair production of Higgs bosons in the $b\bar{b}b\bar{b}$ final state using $pp$ collisions at $\sqrt{s}$ = 13 TeV with the ATLAS detector}",
    eprint = "2202.07288",
    archivePrefix = "arXiv",
    primaryClass = "hep-ex",
    reportNumber = "CERN-EP-2021-229",
    doi = "10.1103/PhysRevD.105.092002",
    journal = "Phys. Rev. D",
    volume = "105",
    number = "9",
    pages = "092002",
    year = "2022"
}

@article{ATLAS:2019qdc,
    author = "Aad, Georges and others",
    collaboration = "ATLAS",
    title = "{Combination of searches for Higgs boson pairs in $pp$ collisions at $\sqrt{s} = $13 TeV with the ATLAS detector}",
    eprint = "1906.02025",
    archivePrefix = "arXiv",
    primaryClass = "hep-ex",
    reportNumber = "CERN-EP-2019-099",
    doi = "10.1016/j.physletb.2019.135103",
    journal = "Phys. Lett. B",
    volume = "800",
    pages = "135103",
    year = "2020"
}

@article{CMS:2022gjd,
    author = "Tumasyan, Armen and others",
    collaboration = "CMS",
    title = "{Search for Nonresonant Pair Production of Highly Energetic Higgs Bosons Decaying to Bottom Quarks}",
    eprint = "2205.06667",
    archivePrefix = "arXiv",
    primaryClass = "hep-ex",
    reportNumber = "CMS-B2G-22-003, CERN-EP-2022-090",
    doi = "10.1103/PhysRevLett.131.041803",
    journal = "Phys. Rev. Lett.",
    volume = "131",
    number = "4",
    pages = "041803",
    year = "2023"
}

@article{CMS:2020tkr,
    author = "Sirunyan, Albert M and others",
    collaboration = "CMS",
    title = "{Search for nonresonant Higgs boson pair production in final states with two bottom quarks and two photons in proton-proton collisions at $ \sqrt{s} $ = 13 TeV}",
    eprint = "2011.12373",
    archivePrefix = "arXiv",
    primaryClass = "hep-ex",
    reportNumber = "CMS-HIG-19-018, CERN-EP-2020-222",
    doi = "10.1007/JHEP03(2021)257",
    journal = "JHEP",
    volume = "03",
    pages = "257",
    year = "2021"
}

@article{CMS:2025qit,
    author = "Hayrapetyan, Aram and others",
    collaboration = "CMS",
    title = "{Search for a new scalar resonance decaying to a Higgs boson and another new scalar particle in the final state with two bottom quarks and two photons in proton-proton collisions at $\sqrt{s}=13$ TeV}",
    eprint = "2508.11494",
    archivePrefix = "arXiv",
    primaryClass = "hep-ex",
    reportNumber = "CMS-B2G-24-001, CERN-EP-2025-160",
    doi = "10.1007/JHEP12(2025)178",
    journal = "JHEP",
    volume = "12",
    pages = "178",
    year = "2025"
}

@article{ATLAS:2025eii,
    author = "Aad, Georges and others",
    collaboration = "ATLAS, CMS",
    title = "{Highlights of the HL-LHC physics projections by ATLAS and CMS}",
    eprint = "2504.00672",
    archivePrefix = "arXiv",
    primaryClass = "hep-ex",
    reportNumber = "ATL-PHYS-PUB-2025-018, CMS-HIG-25-002",
    month = "4",
    year = "2025"
}

@techreport{Collaboration:2928096,
      author        = "CMS Collaboration",
      collaboration = "CMS",
      title         = "{Projection of CMS experimental reach on HH production at
                       HL-LHC}",
      institution   = "CERN",
      reportNumber  = "CMS-NOTE-2025-006, CERN-CMS-NOTE-2025-006",
      address       = "Geneva",
      year          = "2025",
      url           = "https://cds.cern.ch/record/2928096",
}

@article{Chanowitz:1985hj,
    author = "Chanowitz, Michael S. and Gaillard, Mary K.",
    title = "{The TeV Physics of Strongly Interacting W's and Z's}",
    reportNumber = "LBL-19470, UCB-PTH-85/19",
    doi = "10.1016/0550-3213(85)90580-2",
    journal = "Nucl. Phys. B",
    volume = "261",
    pages = "379--431",
    year = "1985"
}

@article{Veltman:1989ud,
    author = "Veltman, Helene G. J.",
    title = "{The Equivalence Theorem}",
    reportNumber = "LBL-27648, UCB-PTH-89/19",
    doi = "10.1103/PhysRevD.41.2294",
    journal = "Phys. Rev. D",
    volume = "41",
    pages = "2294",
    year = "1990"
}

@article{deFlorian:2016spz,
    author = "de Florian, D. and others",
    collaboration = "LHC Higgs Cross Section Working Group",
    title = "{Handbook of LHC Higgs Cross Sections: 4. Deciphering the Nature of the Higgs Sector}",
    eprint = "1610.07922",
    archivePrefix = "arXiv",
    primaryClass = "hep-ph",
    reportNumber = "CERN-2017-002-M, CERN-2017-002",
    doi = "10.23731/CYRM-2017-002",
    journal = "CERN Yellow Rep. Monogr.",
    volume = "2",
    pages = "1--869",
    year = "2017"
}

@techreport{CMS-PAS-HIG-24-015,
      collaboration = "CMS",
      title         = "{Search for triple Higgs production in Run2 data of CMS
                       using 4b2gamma final state.}",
      institution   = "CERN",
      reportNumber  = "CMS-PAS-HIG-24-015",
      address       = "Geneva",
      year          = "2025",
      url           = "https://cds.cern.ch/record/2937680",
}

@article{Papaefstathiou:2020lyp,
    author = "Papaefstathiou, Andreas and Robens, Tania and Tetlalmatzi-Xolocotzi, Gilberto",
    title = "{Triple Higgs Boson Production at the Large Hadron Collider with Two Real Singlet Scalars}",
    eprint = "2101.00037",
    archivePrefix = "arXiv",
    primaryClass = "hep-ph",
    reportNumber = "SI-HEP-2020-34, RBI-ThPhys-2020-53",
    doi = "10.1007/JHEP05(2021)193",
    journal = "JHEP",
    volume = "05",
    pages = "193",
    year = "2021"
}

@article{Carena:2015moc,
    author = "Carena, Marcela and Haber, Howard E. and Low, Ian and Shah, Nausheen R. and Wagner, Carlos E. M.",
    title = "{Alignment limit of the NMSSM Higgs sector}",
    eprint = "1510.09137",
    archivePrefix = "arXiv",
    primaryClass = "hep-ph",
    reportNumber = "FERMILAB-PUB-15-407-T, EFI-15-32, MCTP-15-15, SCIPP-15-12, WSU-HEP-1505",
    doi = "10.1103/PhysRevD.93.035013",
    journal = "Phys. Rev. D",
    volume = "93",
    number = "3",
    pages = "035013",
    year = "2016"
}

@article{Abouabid:2024gms,
    author = "Abouabid, Hamza and others",
    title = "{HHH whitepaper}",
    eprint = "2407.03015",
    archivePrefix = "arXiv",
    primaryClass = "hep-ph",
    doi = "10.1140/epjc/s10052-024-13376-3",
    journal = "Eur. Phys. J. C",
    volume = "84",
    pages = "1183",
    year = "2024"
}

@inproceedings{Butterworth:2008tr,
    author = "Butterworth, Jonathan M. and Davison, Adam R. and Rubin, Mathieu and Salam, Gavin P.",
    title = "{Jet substructure as a new Higgs search channel at the LHC}",
    booktitle = "{34th International Conference on High Energy Physics}",
    eprint = "0810.0409",
    archivePrefix = "arXiv",
    primaryClass = "hep-ph",
    month = "10",
    year = "2008"
}

@article{Larkoski:2017jix,
    author = "Larkoski, Andrew J. and Moult, Ian and Nachman, Benjamin",
    title = "{Jet Substructure at the Large Hadron Collider: A Review of Recent Advances in Theory and Machine Learning}",
    eprint = "1709.04464",
    archivePrefix = "arXiv",
    primaryClass = "hep-ph",
    doi = "10.1016/j.physrep.2019.11.001",
    journal = "Phys. Rept.",
    volume = "841",
    pages = "1--63",
    year = "2020"
}

@article{Radovic:2018dip,
    author = "Radovic, Alexander and Williams, Mike and Rousseau, David and Kagan, Michael and Bonacorsi, Daniele and Himmel, Alexander and Aurisano, Adam and Terao, Kazuhiro and Wongjirad, Taritree",
    title = "{Machine learning at the energy and intensity frontiers of particle physics}",
    reportNumber = "FERMILAB-PUB-18-436-ND",
    doi = "10.1038/s41586-018-0361-2",
    journal = "Nature",
    volume = "560",
    number = "7716",
    pages = "41--48",
    year = "2018"
}

@article{Li:2026kqk,
    author = "Li, Peiran and Liu, Zhen and Wang, Lian-Tao",
    title = "{A Busy Higgs Signal}",
    eprint = "2604.14284",
    archivePrefix = "arXiv",
    primaryClass = "hep-ph",
    month = "4",
    year = "2026"
}

@article{CMS:2024qdd,
    author = "Hayrapetyan, Aram and others",
    collaboration = "CMS",
    title = "{Search for production of a single vectorlike quark decaying to tH or tZ in the all-hadronic final state in pp collisions at s=13{\,}{\,}TeV}",
    eprint = "2405.05071",
    archivePrefix = "arXiv",
    primaryClass = "hep-ex",
    reportNumber = "CMS-B2G-19-001, CERN-EP-2024-067",
    doi = "10.1103/PhysRevD.110.072012",
    journal = "Phys. Rev. D",
    volume = "110",
    number = "7",
    pages = "072012",
    year = "2024"
}

@article{ATLAS:2024mrr,
    author = "Aad, Georges and others",
    collaboration = "ATLAS",
    title = "{Search for vector-like leptons coupling to first- and second-generation Standard Model leptons in pp collisions at $ \sqrt{s} $ = 13 TeV with the ATLAS detector}",
    eprint = "2411.07143",
    archivePrefix = "arXiv",
    primaryClass = "hep-ex",
    reportNumber = "CERN-EP-2024-296",
    doi = "10.1007/JHEP05(2025)075",
    journal = "JHEP",
    volume = "05",
    pages = "075",
    year = "2025"
}

@article{Benbrik:2024fku,
    author = "Benbrik, Rachid and Boukidi, Mohammed and Ech-chaouy, Mohamed and Moretti, Stefano and Salime, Khawla and Yan, Qi-Shu",
    title = "{Vector-Like Quarks at the LHC: A unified perspective from ATLAS and CMS exclusion limits}",
    eprint = "2412.01761",
    archivePrefix = "arXiv",
    primaryClass = "hep-ph",
    doi = "10.1007/JHEP03(2025)020",
    journal = "JHEP",
    volume = "03",
    pages = "020",
    year = "2025"
}

@article{OropezaBarrera:2024ddv,
    author = "Oropeza Barrera, Cristina",
    collaboration = "CMS",
    title = "{Searches for vector-like quarks in CMS}",
    reportNumber = "CMS-CR-2024-275",
    doi = "10.22323/1.478.0011",
    journal = "PoS",
    volume = "LHCP2024",
    pages = "011",
    year = "2025"
}

\end{document}